\documentclass[pra,aps,onecolumn,superscriptaddress,notitlepage,10pt]{revtex4-1}

\pdfoutput=1

\usepackage{graphicx}
\usepackage{amsmath}
\usepackage{amssymb}
\usepackage{times}
\usepackage{color}
\usepackage{bbm}
\usepackage{bm}
\usepackage{soul}

\newcommand{\Eins}{\ensuremath{\mathbbm 1}}
\newcommand{\tr}{{\rm Tr}}
\newcommand{\vect}[1]{\bm{#1}}

\newcommand{\be}{\begin{equation}}
\newcommand{\ee}{\end{equation}}

\newcommand{\beq}{\begin{eqnarray}}
\newcommand{\eeq}{\end{eqnarray}}

\begin{document}

\title{Quantum-enhanced multiparameter estimation in multiarm interferometers}

\author{Mario A. Ciampini}
\affiliation{Dipartimento di Fisica, Sapienza Universit\`{a} di Roma,
Piazzale Aldo Moro 5, I-00185 Roma, Italy}

\author{Nicol\`o Spagnolo}
\email{nicolo.spagnolo@uniroma1.it}
\affiliation{Dipartimento di Fisica, Sapienza Universit\`{a} di Roma,
Piazzale Aldo Moro 5, I-00185 Roma, Italy}

\author{Chiara Vitelli}
\affiliation{Dipartimento di Fisica, Sapienza Universit\`{a} di Roma,
Piazzale Aldo Moro 5, I-00185 Roma, Italy}

\author{Luca Pezz\`e}
\affiliation{QSTAR, INO-CNR and LENS, Largo Enrico Fermi 2, 50125 Firenze, Italy}

\author{Augusto Smerzi}
\affiliation{QSTAR, INO-CNR and LENS, Largo Enrico Fermi 2, 50125 Firenze, Italy}

\author{Fabio Sciarrino}
\email{fabio.sciarrino@uniroma1.it}
\affiliation{Dipartimento di Fisica, Sapienza Universit\`{a} di Roma,
Piazzale Aldo Moro 5, I-00185 Roma, Italy}

\begin{abstract}
Quantum metrology is the state-of-the-art measurement technology. It uses quantum resources to enhance the sensitivity of phase estimation beyond what reachable within classical physics. While single parameter estimation theory has been widely investigated, much less is known about the simultaneous estimation of multiple phases, which finds key applications in imaging and sensing. In this manuscript we provide conditions of useful entanglement (among multimode particles, qudits) for multiphase estimation and adapt them to multiarm Mach-Zehnder interferometry. We discuss benchmark multimode Fock states containing useful qudit entanglement and overcoming the sensitivity of separable qudit states in three and four arm Mach-Zehnder-like interferometers - currently within the reach of integrated photonics technology.
\end{abstract}

\flushbottom
\maketitle

\thispagestyle{empty}

\section*{Introduction}

Quantum metrology exploits entanglement in the probe state to enhance
the precision of parameter estimation beyond what is reachable with classical resources
(see \cite{art:Giovannetti2011, PezzeVarenna2014} for reviews).
The role of entanglement in the estimation of a single parameter
has been clarified~\cite{PhysRevLett.96.010401, PezzePRL2009, HyllusPRA2012, TothPRA2012} and investigated
experimentally in Mach-Zehnder interferometers (MZI)~\cite{art:Krischek2011}.
However, much less is known about the role of entanglement in the joint estimation of multiple parameters.
Yet, in many practical applications, multiple parameters are estimated simultaneously.
This includes quantum imaging \cite{art:Preza} as well as probing of biological samples.
Interestingly, the theory of multiphase estimation does not follow trivially from what is
known about the single parameter case \cite{Hels76}.
Indeed, ultimate multiphase estimation bounds are not saturable in general \cite{MatsumotoJPA1998}, due to
the non-commutativity of the operators generating the phase shift transformations \cite{art:Helstrom1974, art:Yuen1973}.
First insights on this scenario have been recently reported
\cite{MonrasPRA2011,art:spagnolo3d,art:Humphreys,art:Genoni2013,art:Crowley,art:Vidrighin}.

A natural platform for multiparameter quantum metrology is provided by multiport interferometry,
generalizing conventional two-mode interferometry.
Recent progresses in the realization of multiport devices have been achieved by
exploiting integrated photonics \cite{art:Reck1994,art:Nolte2003,art:Kowalevicz2005,Liu06,Poli08,Matt09,Cres10,Metc12,Mean12}.
Three- and four-port beam-splitters (tritters and quarters) have been produced
with integrated optics \cite{art:Suzuki2006,Peru11,Mean12,art:Spagnolo2013}.
This paves the way toward the realization of multiarm interferometers
created by two tritters (quarters) in succession  \cite{art:GregorWeihs1996}.
Quantum-enhanced  single parameter
estimation in  integrated interferometers has been theoretically predicted \cite{art:spagnolo3d}.

In this manuscript we provide conditions of useful entanglement (among multimode particles, qudits)
for the estimation of multiple phases and adapt them to the case of multiarm MZI.
We analyze the simultaneous estimation of multiple phase shifts in an
experimentally relevant framework, with multiphoton Fock states as probe and photon counting measurement.
Our analysis generalizes the case of twin-Fock MZI which has attracted
large experimental \cite{art:Krischek2011, Naga07a, XiangNATPHOT2011, Kacp10} and
theoretical \cite{HollandPRL1993,KimPRA1998,PezzePRL2013} interest for quantum-enhanced single phase estimation.
From the analysis of the Fisher information and employing an adaptive multiphase estimation,
we predict a multiparameter estimation sensitivity beyond the limit achievable with separable qudit probe states.

\section*{Results}

\subsection*{Multiparameter estimation}

\begin{figure}[b!]
\centering
\includegraphics[width=0.49\textwidth]{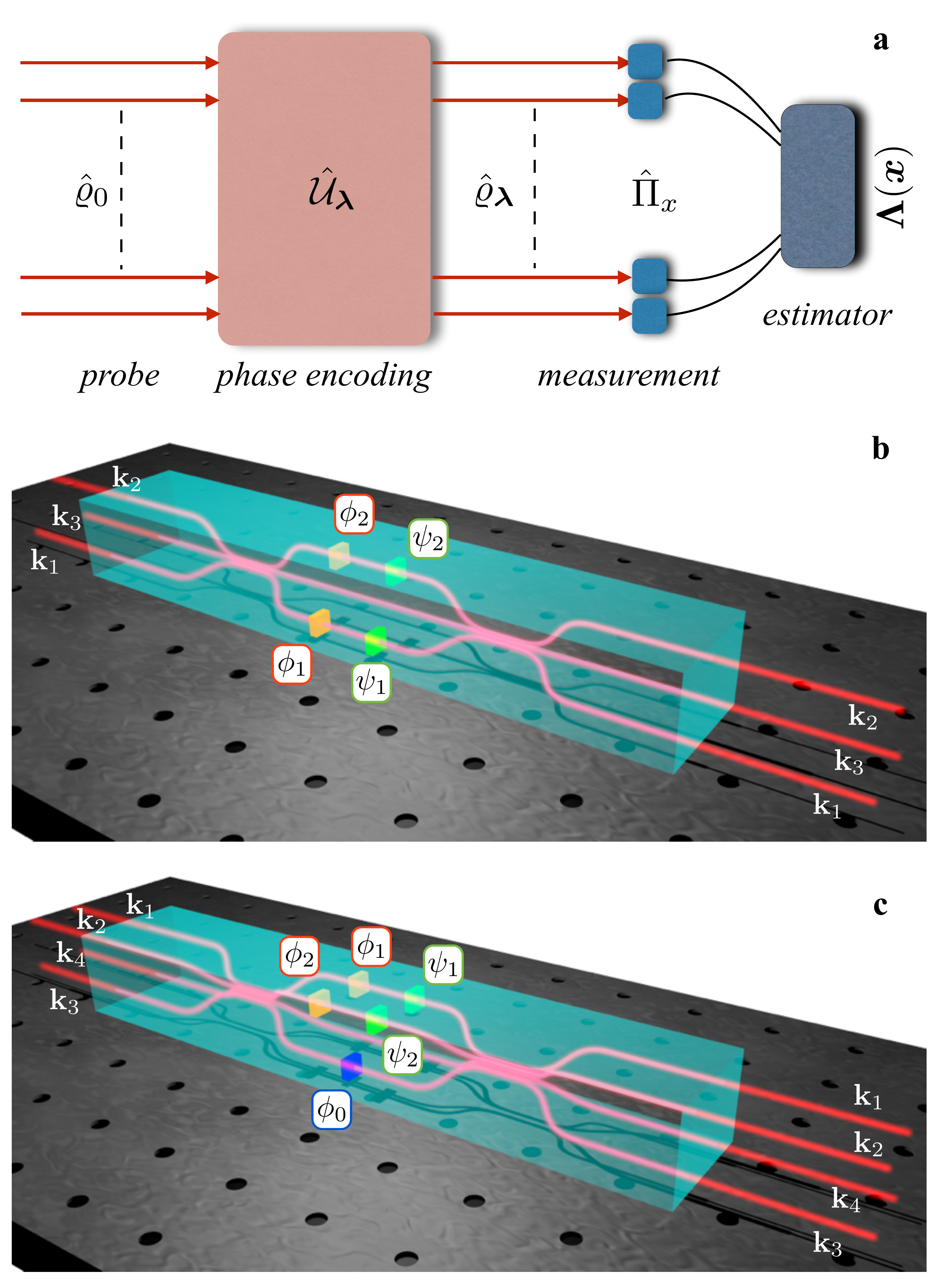}
\caption{\textbf{a.} Conceptual scheme of multiparameter estimation.
\textbf{b.} Three-mode MZI for two-parameter phase estimation built by two cascaded three-port beam-splitters. Phases $(\phi_{1},\phi_{2})$ on modes $(\mathbf{k}_{1},\mathbf{k}_{2})$ are the parameters to be estimated, while $(\psi_{1},\psi_{2})$ are two additional controlled phase-shifts
\textbf{c.} Four-mode interferometer for two-parameter phase estimation built by two cascaded four-port beam-splitters. Phases $(\phi_{1},\phi_{2})$ on modes $(\mathbf{k}_{1},\mathbf{k}_{2})$ are the parameters to be estimated, while ($\phi_0,\psi_{1},\psi_{2}$) are assumed known and controlled.
}
\label{fig:integrated}
\end{figure}

 We consider here the estimation of a $n$-dimensional vector parameter
${\bm \lambda} = (\lambda_{1}, \cdots, \lambda_{n})$ \cite{Hels76}.
In our benchmark, every parameter corresponds to a phase to be estimated in a multiarm interferometer.
A general approach (see Fig.~\ref{fig:integrated} \textbf{a}) consists in  preparing a probe state $\hat{\varrho}_{0}$,
applying a ${\bm \lambda}$-dependent unitary transformation $\hat{\mathcal{U}}_{{\bm \lambda}}$
and performing independent measurements on $\nu$ identical copies of the output state
$\hat{\varrho}_{{\bm \lambda}}= \hat{\mathcal{U}}_{{\bm \lambda}} \hat{\varrho}_{0} \hat{\mathcal{U}}^{\dag}_{{\bm \lambda}}$.
The measurement is described by a positive-operator valued measure (POVM), i.e. a set $\{ \hat{\Pi}_{x} \}$
of positive operators satisfying $\sum_x \hat{\Pi}_x = \Eins$, $P(x\vert \bm{\lambda})={\rm Tr}[\varrho_{{\bm \lambda}} \hat{\Pi}_x]$
being the probability of the detection event $x$.
Finally, the sequence $\vect{x}\equiv(x_1,\cdots,x_\nu)$ of $\nu$ measurement results
is mapped into a vector parameter
${\bm \Lambda}(\vect{x}) = (\Lambda_{1}(\vect{x}), \cdots, \Lambda_{n}(\vect{x}))$,
representing our estimate of $\vect{\lambda}$.
A figure of merit of multiparameter estimation is the covariance matrix
\be
\mathbf{C}_{i,j} = \sum_{\vect{x}}  P(\vect{x} \vert \bm{\lambda})
\big[ \bar{\vect{\Lambda}}_i - \Lambda_i(\vect{x}) \big]
\big[ \bar{\vect{\Lambda}}_j - \Lambda_j(\vect{x}) \big],
\ee
where $P(\vect{x} \vert \bm{\lambda})=\prod_{i=1}^{\nu} P(x_i\vert \bm{\lambda})$ and
$\bar{\vect{\Lambda}} \equiv (\bar{\vect{\Lambda}}_1,\cdots,\bar{\vect{\Lambda}}_n) $ is the mean value of the estimator vector.
For locally unbiased estimators (i.e. $\partial \bar{\Lambda}_i/\partial \lambda_j = \delta_{i,j}$)
the covariance matrix is bounded, via the Cramer-Rao theorem \cite{Hels76}, as
\be \label{CR}
\mathbf{C} \geq \mathbf{F^{-1}}/\nu
\ee
(in the sense of matrix inequality), where
\begin{equation}
\mathbf{F}_{i,j} =
\sum_{x} \frac{1}{p(x \vert {\bm \lambda})} \frac{\partial p(x \vert {\bm \lambda})}{\partial  \lambda_{i}} \frac{\partial p(x \vert {\bm \lambda})}{\partial  \lambda_{j}}
\end{equation}
is the Fisher information matrix (FIM).
Notice that Eq.~(\ref{CR}) can be derived only when the FIM is invertible.
The equality sign in Eq.~(\ref{CR}) is saturated, asymptotically in $\nu$, by the maximum likelihood estimator.
In this Letter we quantify the phase sensitivity by the variance of each estimator,
$(\delta \lambda_{j})^2 \equiv \vect{C}_{j,j}$. We have
\begin{equation} \label{eq:FIM}
(\delta \lambda_{j})^2 \geq \frac{ [\mathbf{F^{-1}}]_{j,j} }{\nu} \geq \frac{1}{\nu \mathbf{F}_{j,j}},
\end{equation}
where the first inequality is due to~(\ref{CR})
and the second follows from a Cauchy-Schwarz inequality, see~\cite{SuppInf}.
Since $1/(\nu \mathbf{F}_{j,j})$ is the Cramer-Rao bound for single parameter estimation,
inequality~(\ref{eq:FIM}) tells us that sensitivity in the estimation of $\lambda_j$
is optimized when fixing the other parameters to known values.
We will also consider
\begin{equation} \label{eq:TrFIM}
\sum_{j=1}^n (\delta \lambda_{j})^{2} \geq \frac{\mathrm{Tr}[{\mathbf{F}}^{-1}]}{\nu}
\geq \frac{1}{\nu} \sum_{j=1}^{n} \frac{1}{\mathbf{F}_{j,j}}.
\end{equation}
The right-hand side inequality in Eqs.~(\ref{eq:FIM}) and~(\ref{eq:TrFIM}) is saturated if and only if the FIM is diagonal,
$\mathbf{F}_{i,j} = F_{i} \delta_{i,j}$.
Furthermore the FIM is bounded by the quantum Fisher information matrix (QFIM): $\mathbf{F} \leq \mathbf{F_Q}$, where
\begin{equation}
[\mathbf{F_Q}]_{i,j} = \mathrm{Tr} \big[ \varrho_{{\bm \lambda}} \hat{L}_{i} \hat{L}_{j} + \varrho_{{\bm \lambda}} \hat{L}_{j} \hat{L}_{i} \big]/2,
\end{equation}
and $\hat{L}_{j}$ is the symmetric logarithmic derivative of $\varrho_{{\bm  \lambda}}$ with respect to parameter $\lambda_{j}$,
defined as $\partial_{j} \varrho_{{\bm \lambda}} = (\hat{L}_{j} \varrho_{{\bm \lambda}}+\varrho_{{\bm \lambda}} \hat{L}_{j})/2$ \cite{Hels76}.
In the single parameter case, the QFIM reduces to a single scalar quantity and it is always possible to find a
POVM for which $F = F_Q$ holds \cite{BraunsteinPRL1994} and $\delta \lambda = 1/F_Q$ holds.
In contrast, in the general multiparameter case, it is not possible to achieve this bound \cite{MatsumotoJPA1998,art:Helstrom1974, art:Yuen1973}.

\subsection*{Sensitivity bounds for separable states}

Here we set sensitivity bounds for multiparameter estimation
when the probe state is separable.
A state $\hat{\varrho}_{0}$ of $N$ qudits is said to be qudit-separable if it can be written as
$\hat{\varrho}_{\rm sep} = \sum_k p_k \hat{\varrho}_k^{(1)} \otimes \cdots \otimes \hat{\varrho}_k^{(N)}$,
where $\hat{\varrho}_j^{(k)}$ $(j=1,\cdots,N)$ is a single qudit state.
A state which is not qudit-separable is qudit-entangled.
The notion of qudit (a particle in $n>2$ modes) generalizes
the concept of qubit (a two-mode particle) and is relevant
in multimode interferometry \cite{PezzeVarenna2014}.
We will set the conditions of qudit entanglement when the generator of each phase shift,
$\hat{G}_j \equiv i \frac{\partial \hat{\mathcal{U}}_{{\bm \lambda}}}{\partial \lambda_j} \hat{\mathcal{U}}^{\dag}_{{\bm \lambda}}$,
is local in the qudit, {\it i.e.} it can be written as $\hat{G}_j = \sum_{i=1}^N \hat{g}_j^{(i)}$
where $\hat{g}_j^{(i)}$ acts on the $i$th qudit.
For simplicity, we will take the same operator $\hat{g}_j^{(i)}= \hat{g}_j$
for each particle.
For separable probe states the inequality
\be \label{eq:Fineq}
\mathbf{F}_{j,j} \leq N ( g_{j,\max}-g_{j,\min} )^2
\ee
holds for all possible POVMs~\cite{SuppInf}, where $g_{j,\max}$ and $g_{j,\min}$ are
the maximum and minimum eigenvalue of $\hat{g}_j$, respectively.
Inequality~(\ref{eq:Fineq}) gives a bound on the sensitivity reachable with separable states
for the estimation of the single parameter $\lambda_j$, when all other parameters are set to a known value.
Inequality~(\ref{eq:Fineq}) can be always saturated by optimal measurements~\cite{SuppInf}.
For the estimation of the single parameter, the violation of Eq.~(\ref{eq:Fineq})
is a necessary and sufficient condition of useful qudit entanglement~\cite{PezzeVarenna2014, PezzePRL2009}.
Regarding the simultaneous estimation of multiple parameters,
we can use Eq.~(\ref{eq:Fineq}) and the chain of inequalities~(\ref{eq:FIM}) to obtain
\be \label{eq:sep}
[{\mathbf{F}}^{-1}]_{j,j} \geq \frac{1}{ N ( g_{j,\max}-g_{j,\min} )^2 }.
\ee
Inequality (\ref{eq:sep}) is a bound of sensitivity in the estimation of the
single parameter $\lambda_{j}$ with separable states, when all the parameters are unknown.
Summing Eq.~(\ref{eq:sep}) over all parameters, we obtain
\be \label{eq:Trsep}
\mathrm{Tr}[{\mathbf{F}}^{-1}] \geq \frac{1}{N} \sum_{j=1}^n \frac{1}{ ( g_{j,\max}-g_{j,\min} )^2 }.
\ee
According to Eqs.~(\ref{eq:sep}) and (\ref{eq:Trsep}), for separable states such that the FIM is invertible, we recover -- at best -- the shot noise scaling
of phase sensitivity, $\delta \lambda_{j} \propto N^{-1/2}$, which characterizes single parameter estimation.
The optimal prefactor, which is equal to one in the qubit case, reflects the multimode nature of the problem.
%%%%%%%%%%%%%%%%%%%%%%%%%%%%%%%%%%%%%%%%%%%%%%%%%%%%%%%%%%%%%%%
% figure 1
%%%%%%%%%%%%%%%%%%%%%%%%%%%%%%%%%%%%%%%%%%%%%%%%%%%%%%%%%%%%%%%
%%%%%%%%%%%%%%%%%%%%%%%%%%%%%%%%%%%%%%%%%%%%%%%%%%%%%%%%%%%%%%%
%%%%%%%%%%%%%%%%%%%%%%%%%%%%%%%%%%%%%%%%%%%%%%%%%%%%%%%%%%%%%%%
%%%%%%%%%%%%%%%%%%%%%%%%%%%%%%%%%%%%%%%%%%%%%%%%%%%%%%%%%%%%%%%

%%%%%%%%%%%%%%%%%%%%%%%%%%%%%%%%%%%%%%%%%%%%%%%%%%%%%%%%%%%%%%%
% figure 2
%%%%%%%%%%%%%%%%%%%%%%%%%%%%%%%%%%%%%%%%%%%%%%%%%%%%%%%%%%%%%%%
\begin{figure*}[ht!]
\centering
\includegraphics[width=0.99\textwidth]{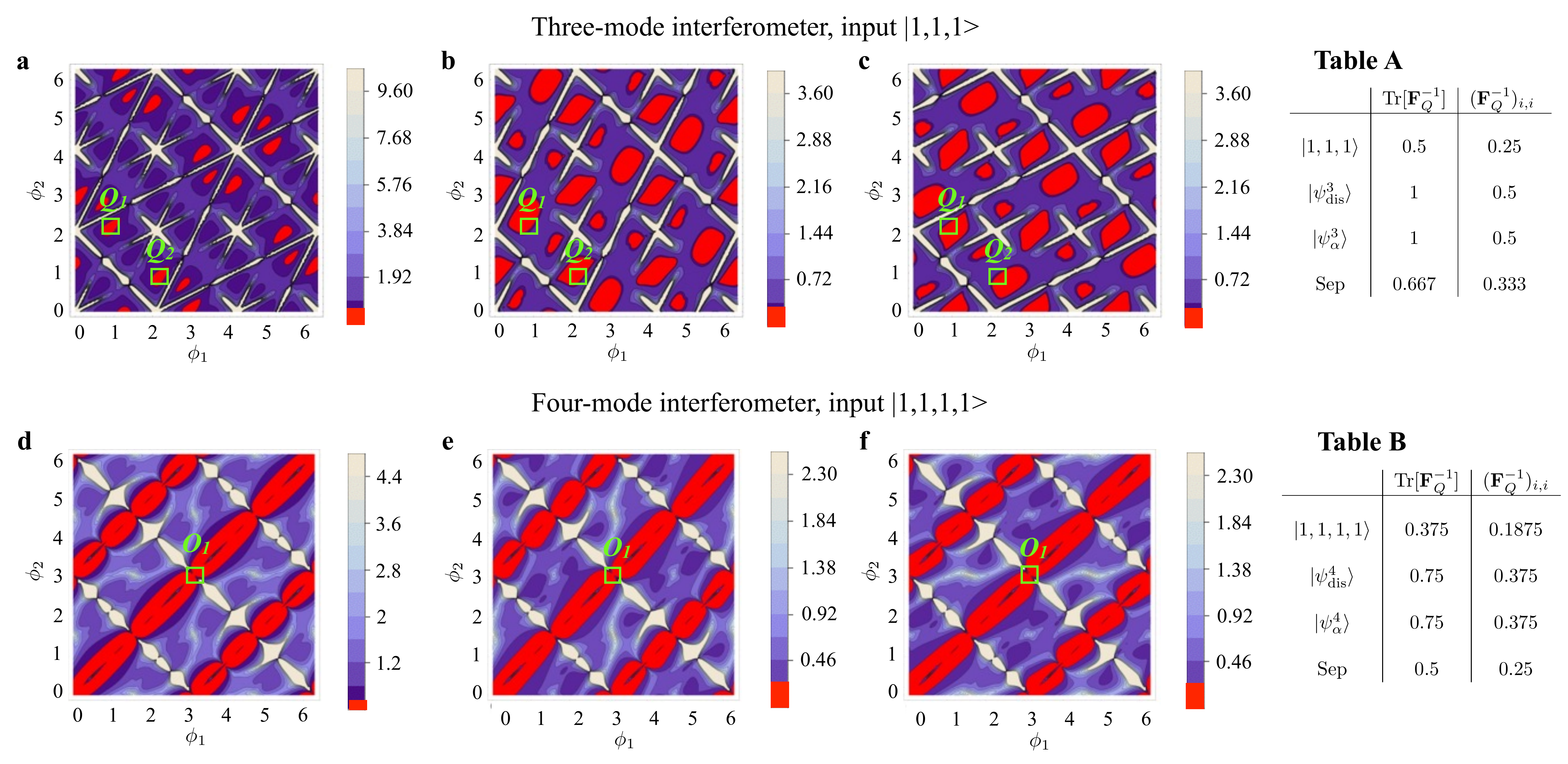}
\caption{\textbf{a-c.} Optimal phase sensitivity of the three-mode balanced MZI with $\vert 1,1,1 \rangle$ probe state  and photon-number measurement. Contour plots of \textbf{a.} $\mathrm{Tr}[\mathbf{F}^{-1}]$, \textbf{b.} $(\mathbf{F}^{-1})_{1,1}$, \textbf{c.} $(\mathbf{F}^{-1})_{2,2}$, as a function of $\phi_{1}$ and $\phi_{2}$. $\mathrm{Tr}[\mathbf{F}^{-1}]$ is minimized at the working points $Q_{1}$ and $Q_{2}$ (see main text).
\textbf{d-f.} Optimal phase sensitivity of the four-mode balanced MZI with $\vert 1,1,1,1 \rangle$ probe state and photon-number measurement. Contour plots of \textbf{d.} $\mathrm{Tr}[\mathbf{F}^{-1}]$, \textbf{e.} $(\mathbf{F}^{-1})_{1,1}$, \textbf{f.} $(\mathbf{F}^{-1})_{2,2}$, as a function of $\phi_{1}$ and $\phi_{2}$. These are shown for $\phi_0=0.001$ to avoid undetermined points in the plot. The QCRB is achieved, for instance, at the point $O_1=[\pi,\pi]$.
Red areas indicate the violation of the separable bound defined in Eq. (\ref{eq:MMZIsep}).
Tables {\bf A} and {\bf B} report $\mathrm{Tr}[\mathbf{F}_{Q}^{-1}]$ and $(\mathbf{F}_{Q}^{-1})_{i,i}$ for different input states and their comparison with the separable bound (Sep).
}
\label{fig:CFI_Fock}
\end{figure*}
%%%%%%%%%%%%%%%%%%%%%%%%%%%%%%%%%%%%%%%%%%%%%%%%%%%%%%%%%%%%%%%
%%%%%%%%%%%%%%%%%%%%%%%%%%%%%%%%%%%%%%%%%%%%%%%%%%%%%%%%%%%%%%%
%%%%%%%%%%%%%%%%%%%%%%%%%%%%%%%%%%%%%%%%%%%%%%%%%%%%%%%%%%%%%%%

\subsection*{Multimode Mach-Zehnder interferometry}
In the following we discuss the estimation of a phase vector ${\bm \phi} = (\phi_{1},\ldots,\phi_{n})$
in a multiarm Mach-Zehnder interferometer (MMZI)
(see Figs.~\ref{fig:integrated} \textbf{b}-\textbf{c}).
The MMZI is built by cascading a $d$-mode balanced beam-splitter $\hat{\mathcal{U}}^{(d)}$
-- which is the natural extension of the standard 50-50 beam-splitter to more than two optical input-output modes \cite{book:zeilinger1993} --
a phase shift transformation $\hat{\mathcal{U}}({\bm \phi}) = e^{-\imath \sum_{i=1}^n \hat{N}_{i} \phi_{i}}$, being $\hat{N}_{i}$
the photon-number operator for the $i$th mode,
and a second multiport beam-splitter $\hat{\mathcal{U}}^{(d)}$.
Hence, this platform can be adopted as a benchmark to investigate simultaneous estimation of $n=d-1$
optical phases. Indeed, it allows for a direct comparison between classical and quantum probe states, and can be adapted to represent a flexible
platform for the analysis of multiparameter scenario by changing the unitary transformation of the input and output multiport splitters.
The inequalities~(\ref{eq:sep}) and (\ref{eq:Trsep}) thus read
\be \label{eq:MMZIsep}
[{\mathbf{F}}^{-1}]_{j,j} \geq \frac{1}{N}, \qquad \text{and} \qquad \mathrm{Tr}[{\mathbf{F}}^{-1}] \geq \frac{n}{N},
\ee
respectively.
The recent experimental implementation of symmetric multiport beam-splitter
\cite{Mean12,art:Spagnolo2013}, by adopting integrated platforms, paves the way toward the
future realization of optical MMZI.
For $d=3$ modes, $\hat{\mathcal{U}}^{(3)}$ (tritter) has
diagonal elements $(\hat{\mathcal{U}}^{(3)})_{i,i}=3^{-1/2}$ and off-diagonal elements $(\hat{\mathcal{U}}^{(3)})_{i,j}=3^{-1/2} e^{\imath 2 \pi/3}$ with $i \neq j$.
For $d=4$ modes, $\hat{\mathcal{U}}^{(4)}$ (quarter) is
$(\hat{\mathcal{U}}^{(4)})_{i,i}=2^{-1}$ and $(\hat{\mathcal{U}}^{(4)})_{i,j}=-2^{-1}$ for $i \neq j$.
The phase vector is estimated from the measurement of the number of particles in each mode.

%%%%%%%%%%%%%%%%%%%%%%%%%%%%%%%%%%%%%%%%%%%%%%%%%%%%%%%%%%%%%%%
% figure 3
%%%%%%%%%%%%%%%%%%%%%%%%%%%%%%%%%%%%%%%%%%%%%%%%%%%%%%%%%%%%%%%
\begin{figure}[ht!]
\centering
\includegraphics[width=0.75\textwidth]{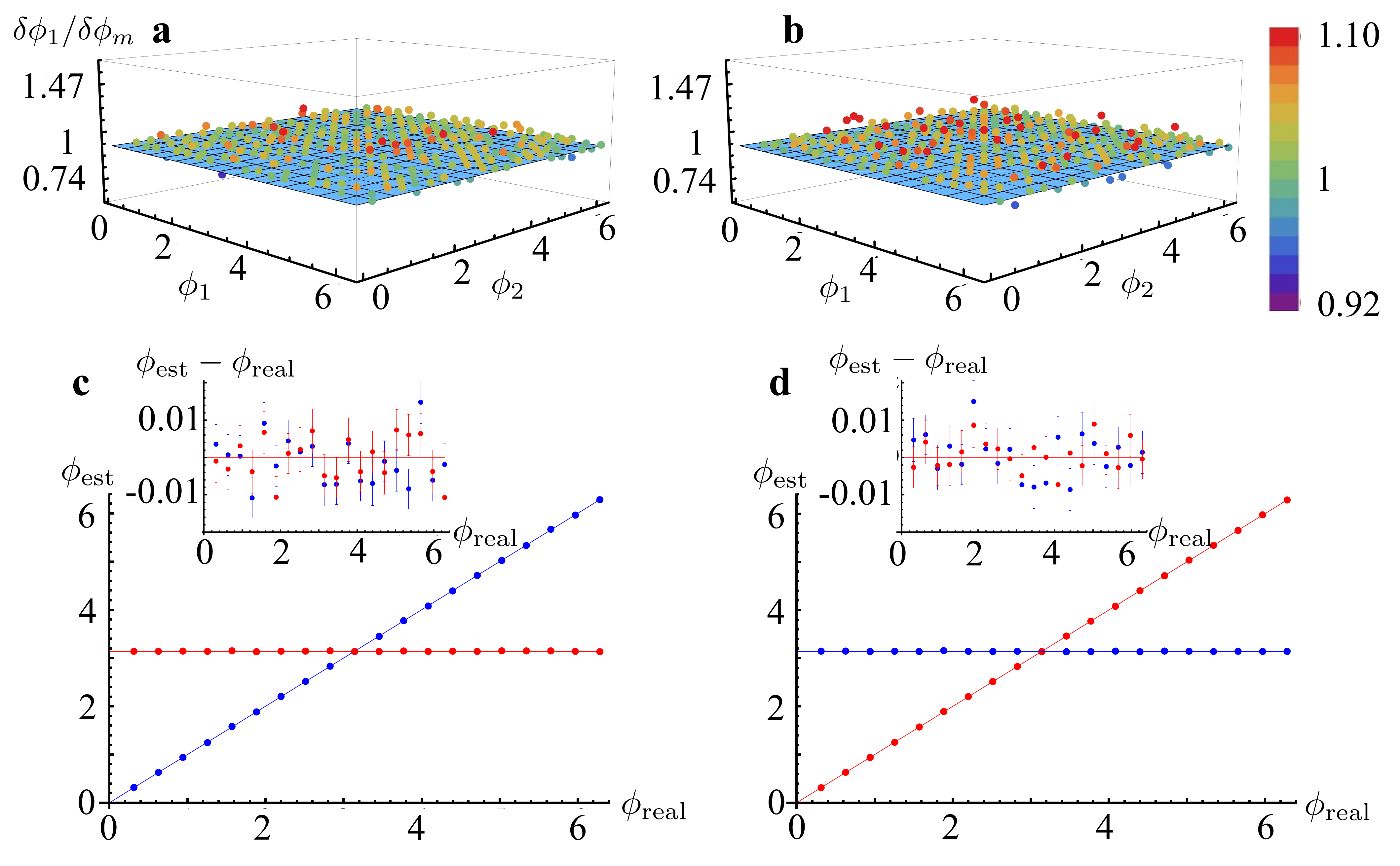}
\caption{Adaptive estimation of two phases, $\phi_{1}$ and $\phi_{2}$ with
the three-mode interferometer injected by a $\vert 1,1,1 \rangle$.
The adaptive protocol (see text) aims at reaching a phase uncertainty $\delta \phi_{1} \approx \delta \phi_2$
after $\nu=10000$ independent measurements.
\textbf{a-b}: Uncertainties $\delta \phi_{1}/\delta \phi_{m} $ and $\delta \phi_{2}/\delta \phi_{m} $ obtained for different values of $\phi_{1}$ and $\phi_{2}$ (points)
and normalized with respect to the expected value $\delta \phi_{m} = 0.543/\sqrt{\nu}$ (see text).
As an example, we report the results obtained for the specific cases $\phi_1 = \pi$ (\textbf{c})
and $\phi_2 = \pi$ (\textbf{d}). In these panels the blue line is the estimated value of $\phi_1$, the red line is the estimated $\phi_2$.
The inset shows the difference between the estimated value and the actual value of the phases, error bars are obtained by repeating $p=1000$ times the numerical simulation of the protocol.
}
\label{fig:Adaptive_Fock_3}
\end{figure}
%%%%%%%%%%%%%%%%%%%%%%%%%%%%%%%%%%%%%%%%%%%%%%%%%%%%%%%%%%%%%%%
%%%%%%%%%%%%%%%%%%%%%%%%%%%%%%%%%%%%%%%%%%%%%%%%%%%%%%%%%%%%%%%
%%%%%%%%%%%%%%%%%%%%%%%%%%%%%%%%%%%%%%%%%%%%%%%%%%%%%%%%%%%%%%%

As probe, we focus on multimode Fock states with a single photon in each mode.
For the three-mode MZI, the probe is $\vert 1,1,1 \rangle$,  corresponding to the injection of a single photon in each input mode of the interferometer.
The results of the calculation for $\mathbf{F}^{-1}$ are shown in Fig. \ref{fig:CFI_Fock} {\bf a-c}.
We observe that $\tr[\mathbf{F^{-1}}]$ and the diagonal elements $[\mathbf{F^{-1}}]_{1,1}$ and $[\mathbf{F^{-1}}]_{2,2}$
strongly depends on the phases $\phi_1$ and $\phi_2$.
Notably, the inequalities (\ref{eq:MMZIsep}) are violated at certain optimal values of the parameters, signaling that
the Fock state $\vert 1,1,1\rangle$ contains useful qudit entanglement.
Additionally, we observe characteristic features.
(i) $\vect{F} \neq \vect{F}_Q$, in particular, the minimum value of $\mathrm{Tr}[\mathbf{F}^{-1}]$ is greater than the corresponding minimum value of the QFIM:
$\min_{\phi_{1},\phi_{2}}\mathrm{Tr}[\mathbf{F}^{-1}] = 0.59 > \mathrm{Tr}[\mathbf{F}_{Q}^{-1}] = 0.5$ (see Fig. \ref{fig:CFI_Fock} \textbf{a}).
This value is lower than the bound for separable states given by Eq. (\ref{eq:MMZIsep}) and equal to $\mathrm{Tr}[\mathbf{F}^{-1}] \geq 0.667$
(here $N=3$ and $n=2$).
We also have $\min_{\phi_{1},\phi_{2}} [\mathbf{F}^{-1}]_{1,1} = 0.25 $ and $\min_{\phi_{1},\phi_{2}} [\mathbf{F}^{-1}]_{2,2} = 0.25$, which are smaller than the
bound 0.33 for separable state (see Fig.~\ref{fig:CFI_Fock} \textbf{b-c}).
(ii) The FIM is not always invertible: at the phase values for which $\det \mathbf{F} = 0$
the bound (\ref{CR}) is not defined. Around these points
(white regions in Figs. \ref{fig:CFI_Fock} \textbf{a-c}) $[\mathbf{F^{-1}}]_{1,1}$ and/or $[\mathbf{F^{-1}}]_{2,2}$ diverges.
(iii) The working points to obtain the minimum of the multiparameter bound do not lead to symmetric errors on the single parameters $\phi_{1}$ and $\phi_{2}$. More specifically, when $\mathrm{Tr}[\mathbf{F}^{-1}] = 0.59$, the bounds for the error on the single parameters are different: $\delta \phi_{1}^{\mathrm{min}} \neq \delta \phi_{2}^{\mathrm{min}}$. This is obtained for instance for working point $Q_{1}=(\phi_{1},\phi_{2})=(0.892,2.190)$, leading to
$([\mathbf{F^{-1}}]_{1,1}, [\mathbf{F^{-1}}]_{2,2}) \simeq (0.282,0.309) $
and for working point $Q_{2}=(\phi_{1},\phi_{2})=(2.190,0.892)$, leading to
$([\mathbf{F^{-1}}]_{1,1}, [\mathbf{F^{-1}}]_{2,2}) \simeq (0.309,0.282) $, see Fig.~\ref{fig:CFI_Fock}\textbf{a}.

In summary, with this choice of system and measurement it is not possible to saturate the quantum Cramer-Rao inequality simultaneously for the two parameters. Furthermore, according to point (iii) an adaptive estimation strategy (which we discuss below) is necessary to obtain the minimum sensitivity on both parameters with symmetric errors, and thus saturate the multiparameter Cramer-Rao bound.

We have repeated the above analysis for a four-mode interferometer ($d=4$) with two unknown phases, $\phi_1$ and $\phi_2$, and a known control phase $\phi_0$ (see Fig \ref{fig:integrated} \textbf{c}). This configuration allows a comparison between three- and four-arm interferometers for the two parameter estimation. In the latter case the control phase $\phi_0$ gives us an additional degree of freedom. We choose as input the Fock State $|1,1,1,1\rangle$. In Fig. \ref{fig:CFI_Fock} {\bf d-f} the results of our calculations are reported for a fixed value of $\phi_0$, as well as the numerical analysis of $\det \mathbf{F}$. We observe that as in the previous cases the FIM depends on the value of the parameter to be estimated.  Furthermore, also in the four-mode the achievtable sensitivity falls below the bound (\ref{eq:MMZIsep}) for separable states: we have
$\min_{\phi_{1},\phi_{2}}\mathrm{Tr}[\mathbf{F}^{-1}] = 0.375 $, $\min_{\phi_{1},\phi_{2}} [\mathbf{F}^{-1}]_{1,1} = 0.1875$ and $\min_{\phi_{1},\phi_{2}} [\mathbf{F}^{-1}]_{2,2} = 0.1875$
which are below the bounds 0.5 and 0.25 given by Eq.~(\ref{eq:MMZIsep}) ($N=4$ and $n=2$, here), respectively.
The most notable difference with respect to the previous case is that the QCRB is achieved, for instance in working point $O_{1}=[\pi,\pi]$. In addition, both diagonal terms are equivalent and only a two step protocol is needed (see discussion below).

We have also compared the obtained results with the one achievable with other probe states. For instance, we consider a set of distinguishable particles $\vert \psi_{\mathrm{dis}}^{d} \rangle = \otimes_{q=1}^{d} \vert q \rangle$ (where $\vert q \rangle$ stands for a single photon on mode $\mathbf{k}_{q}$), or an input coherent state $\vert \psi_{\alpha}^{d} \rangle$ on input mode $\mathbf{k}_{1}$ with $\alpha=\sqrt{3}$ for $d=3$ ($\alpha=2$ for $d=4$) and no phase reference. We obtain $\mathrm{Tr}[\mathbf{F}_{Q}^{-1}] = 1$ for both $\vert \psi_{\mathrm{dis}}^{3} \rangle$ and $\vert \psi_{\alpha}^{3} \rangle$, within the bound $\mathrm{Tr}[\mathbf{F}^{-1}] \geq 0.667$
given by Eq. (\ref{eq:MMZIsep}) for separable inputs. Similarly, $\mathrm{Tr}[\mathbf{F}_{Q}^{-1}] = 0.75$ for both $\vert \psi_{\mathrm{dis}}^{4} \rangle$ and $\vert \psi_{\alpha}^{4} \rangle$, within the bound $\mathrm{Tr}[\mathbf{F}^{-1}] \geq 0.5$. Results are summarized in Table A and B.

\subsection*{Adaptive phase estimation}
The above analysis has shown that the working regime in which the minimum uncertainty for the estimation of the two phases $\phi_{1}$ and $\phi_{2}$
with the three-mode interferometer does not give the same error on the two individual parameters.
To overcome this limitation -- and obtain approximatively a symmetric estimate of the two phases $\phi_{1}$ and $\phi_{2}$ -- we exploited an adaptive algorithm.
The protocol requires $\nu$ independent measurements and the adoption of controlled phase shifts $\psi_{i}$ on modes $\mathbf{k}_{i}$, with $i=1,2$,
which have to be tuned during the protocol to perform the estimation in different working points (see in Fig. \ref{fig:integrated} \textbf{b}).
It is divided in three-steps.
In a first step, we set $\psi_{1,2}=0$ and obtain a rough estimate of the phases $\phi_{i}$
after a number of measurements much smaller than $\nu$.
Then, in step 2 the tunable phases $\psi_{i}$ are adjusted so that $\phi_{i}+\psi_{i}$ on arms 1 and 2 are set to be close to the working point $Q_{1}$. In this step essentially half of the remaining resources are spent so as to obtain $(\phi_1^{(Q_1)}+\psi_1)\pm \delta \phi_{1}^{(Q_1)}$ and $(\phi_2^{(Q_1)}+\psi_2)\pm \delta \phi_{2}^{(Q_1)}$ with an adequate estimator. Here $\phi_i^{(Q_1)}$, $\delta \phi_i^{(Q_1)}$ represent respectively the estimation and the uncertainty of $\phi_i$ around working point $Q_1$.
In step 3 the same procedure is repeated for working point $Q_{2}$. Finally  the tunable phases $\Psi_{1.2}$ are subtracted so to recover $\phi_{1,2}\pm \delta\phi_{1,2}$.

The results are shown in Fig. \ref{fig:Adaptive_Fock_3} {\bf a-d} where $\nu$ independent measurements are performed. Half of the measurements ($\nu_{1}=\nu/2$) are performed in point $Q_{1}$, where $\delta \phi_{1} \simeq 0.531/\sqrt{\nu_{1}}$ and $\delta \phi_{2} \simeq 0.556/\sqrt{\nu_{1}}$, while the other half ($\nu_{2}=\nu/2$) are performed in point $Q_{2}$, where $\delta \phi_{1} \simeq 0.556/\sqrt{\nu_{2}}$ and $\delta \phi_{2} \simeq 0.531/\sqrt{\nu_{2}}$. The expected error on a single phase $\delta \phi_{i}$ after the two steps is then obtained as an appropriate combination of the values on the points $Q_{i}$. More specifically, as the Fisher information is additive, the overall FIM reads ${\bf F} = \nu_{1} {\bf F}_{1} + \nu_{2} {\bf F}_{2}$, where ${\bf F}_{i}$ is the FIM in working points $Q_{i}$. We observe that the protocol permits to achieve the bound of the working point, which for $\nu_{1}=\nu_{2}$ is $\delta \phi_1 = \delta \phi_2 \geq 0.543/\sqrt{\nu}$. Note that the bound is lower than the corresponding limit (\ref{eq:MMZIsep}) for separable states $\delta \phi_{i} \geq 0.577/\sqrt{\nu}$.

\begin{figure}[ht!]
\centering
\includegraphics[width=0.75\textwidth]{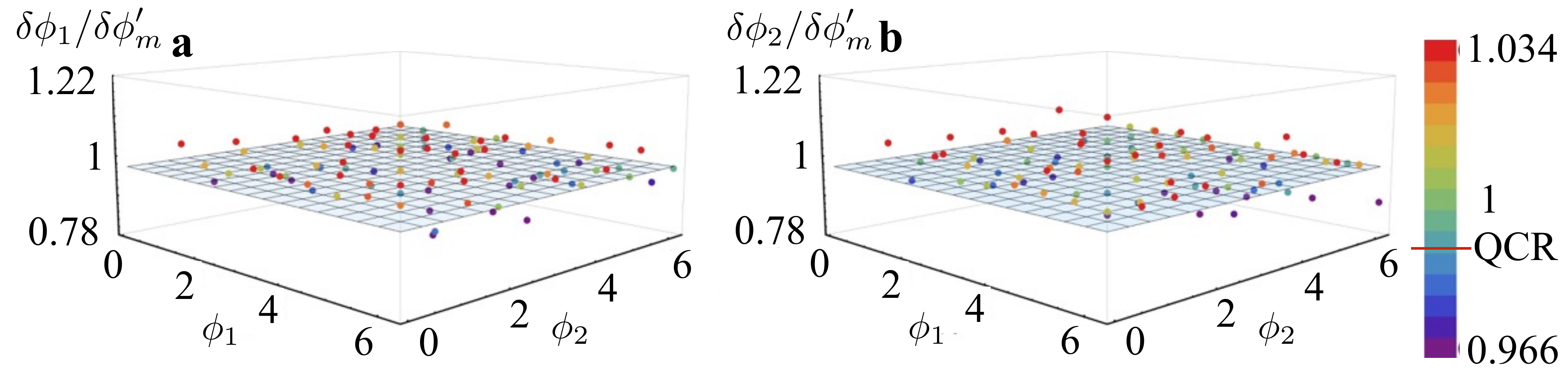}
\caption{Adaptive estimation of two phases, $\phi_{1}$ and $\phi_{2}$ with
the four-mode interferometer injected by a $\vert 1,1,1,1 \rangle$, for $\phi_{0} = 0.01$
and $\nu=10000$ independent measurements.
\textbf{a-b}: Uncertainties $\delta \phi_{1}/\delta \phi_{m}^{\prime}$ and $\delta \phi_{2}/\delta \phi_{m}^{\prime}$ obtained for different values of $\phi_{1}$ and $\phi_{2}$ (points)
and normalized with respect to the achievable bound $\delta \phi_{m}^{\prime} = 0.437/\sqrt{\nu}$. The horizontal red line in the legend corresponds to the quantum Cramer-Rao bound for the single-parameter.}
\label{fig:Adaptive_Fock_4}
\end{figure}

The adaptive scheme for the four-mode interferometer is slightly different:
in this case there is one working point, the point $O_1$, see Fig.~\ref{fig:CFI_Fock}, so we we apply a two-step protocol.
In the first step, we obtain a rough estimate of the parameters with an initial error $\delta$.
Then, in the second step we apply two supplementary phases $\psi_1$ and $\psi_2$ to translate the working point of the protocol to the neighbourhood of $O_{1}$. It should be noticed that a convergent estimation protocol in the second step requires to set $\phi_0$ such that the quantity $\mathrm{Tr}[\mathbf{F}^{-1}]$ has no singularities.
Note that the more $\phi_{0}$ deviates from $\phi_{0}=0$, the larger is the regular region around $O_{1}$ \cite{SuppInf}.
The price to pay is a slightly increasing the error in the estimation process.
The results of the protocol for the four-mode case with $\phi_{0}=0.01$ are then shown in Fig. \ref{fig:Adaptive_Fock_4} {\bf a-b}. Similarly to the three-mode case, we observe that the protocol permits to achieve the bound of the working point, which is $\delta \phi_1 = \delta \phi_2 \geq 0.437/\sqrt{\nu} =  \delta \phi_{m}^{\prime}$ for $\phi_0=0.01$ (plane in Fig. \ref{fig:Adaptive_Fock_4}), while the quantum Cramer-Rao bound reads $\delta \phi_{i} \geq 0.433/\sqrt{\nu}$. This shows that achieving a convergent numerical protocol leads to a slight decrease in phase sensitivity due to singular points in the neighbourhood of the working regions.

%%%%%%%%%%%%%%%%%%%%%%%%%%%%%%%%%%%%%%%%%%%%%%%%%%%%%%%%%%%%%%%
% figure 4
%%%%%%%%%%%%%%%%%%%%%%%%%%%%%%%%%%%%%%%%%%%%%%%%%%%%%%%%%%%%%%%
%%%%%%%%%%%%%%%%%%%%%%%%%%%%%%%%%%%%%%%%%%%%%%%%%%%%%%%%%%%%%%%
%%%%%%%%%%%%%%%%%%%%%%%%%%%%%%%%%%%%%%%%%%%%%%%%%%%%%%%%%%%%%%%
%%%%%%%%%%%%%%%%%%%%%%%%%%%%%%%%%%%%%%%%%%%%%%%%%%%%%%%%%%%%%%%

\section*{Conclusions}
In this manuscript we have developed the general theory of quantum enhanced multi-phase estimation.
In particular, we provide the conditions of useful entanglement among multimode particles (qudits)
for the simultaneous estimation of multiple phases below the ultimate sensitivity limit achievable with separable states. In a realistic experimental scenario, using multi-mode Mach-Zehnder interferometers and photo-counting measurements, Fock state probes can be exploited for multiphase estimation with quantum-enhancement phase sensitivity.
With respect to the estimation of a single phase, where Fock states are known to be a useful resource, our analysis evidences a much richer case:
most notably, the phase sensitivity strongly depends on the
phase value (the Cramer-Rao bound being not always definite)
and on the interferometer configurations  such as the three- and four-mode interferometers.
Finally, we discuss and numerically simulate an adaptive estimation protocol which permits to achieve the expected bounds.
The adaptive strategy becomes crucial in the multiparameter scenario since the simultaneous saturation of the ultimate limits for all parameters is in general not guaranteed.

During the completion of this manuscript, a first implementation of a tritter-based interferometer for single-phase estimation has been reported \cite{Chab14}.

\section*{Acknowledgements}

This work was supported by ERC-Starting Grant 3D-QUEST (3D-Quantum Integrated Optical Simulation; grant agreement no. 307783, http://www.3dquest.eu), EU-STREP Project QIBEC and PRIN project Advanced Quantum Simulation and Metrology (AQUASIM).
LP acknowledges financial support by MIUR through FIRB Project No. RBFR08H058.

\end{document}

\title{Supplementary Information: Quantum-enhanced multiparameter estimation in multiarm interferometers}

\author{Mario A. Ciampini}
\affiliation{Dipartimento di Fisica, Sapienza Universit\`{a} di Roma,
Piazzale Aldo Moro 5, I-00185 Roma, Italy}

\author{Nicol\`o Spagnolo}
\email{nicolo.spagnolo@uniroma1.it}
\affiliation{Dipartimento di Fisica, Sapienza Universit\`{a} di Roma,
Piazzale Aldo Moro 5, I-00185 Roma, Italy}

\author{Chiara Vitelli}
\affiliation{Dipartimento di Fisica, Sapienza Universit\`{a} di Roma,
Piazzale Aldo Moro 5, I-00185 Roma, Italy}

\author{Luca Pezz\`e}
\affiliation{QSTAR, INO-CNR and LENS, Largo Enrico Fermi 2, 50125 Firenze, Italy}

\author{Augusto Smerzi}
\affiliation{QSTAR, INO-CNR and LENS, Largo Enrico Fermi 2, 50125 Firenze, Italy}

\author{Fabio Sciarrino}
\email{fabio.sciarrino@uniroma1.it}
\affiliation{Dipartimento di Fisica, Sapienza Universit\`{a} di Roma,
Piazzale Aldo Moro 5, I-00185 Roma, Italy}

\maketitle

\section{Bounds on the diagonal elements of the Fisher information matrix}

Here we detail the demonstration of the inequality \cite{KayBOOK}
\be \label{Ineq:Fii}
[\mathbf{F}^{-1}]_{i,i} \geq \frac{1}{ {\mathbf{F}}_{i,i} }.
\ee 
To prove Eq.~(\ref{Ineq:Fii}) we recall that the square root of the positive definite Fisher matrix
is given by the matrix with the same (orthonormal) eigenvectors as $\mathbf{F}$ and  
the square root of its eigenvalues.
We indicate as $f_i > 0$ and $\vect{v}_i$ (with $\vect{v}_i^{\top} \vect{v}_j = \delta_{i,j}$)
the eigenvalues and eigenvectors of $\mathbf{F}$, respectively ($\mathbf{F} \vect{v}_i = f_i \vect{v}_i$).
Notice that $\mathbf{F}$ is real and symmetric and thus diagonalize.
In addition, $\mathbf{F}$ is positive semidefinite (therefore $f_i \geq 0$) and 
assuming that $\mathbf{F}$ is invertible we have $f_i\neq 0$. 
We write 
\beq
\mathbf{F} &=& \sum_i f_i \, \vect{v}_i \vect{v}_i^{\top}, \nonumber \\
\sqrt{\mathbf{F}} &=& \sum_i \sqrt{f_i} \, \vect{v}_i \vect{v}_i^{\top}, \nonumber \\
(\sqrt{\mathbf{F}})^{-1} &=& \sqrt{\mathbf{F}^{-1}} = \sum_i \frac{1}{\sqrt{f_i}} \, \vect{v}_i \vect{v}_i^{\top}, \nonumber
\eeq
and therefore
\be
\sqrt{\mathbf{F}}(\sqrt{\mathbf{F}})^{-1} = \sqrt{\mathbf{F}}\sqrt{\mathbf{F}^{-1}} =\Eins, \nonumber
\ee
with $\Eins$ the identity matrix.
Using the Cauchy-Schwarz inequality
\beq
1 &=& \Big(\sum_k [\sqrt{\mathbf{F}}]_{i,k} [\sqrt{\mathbf{F}^{-1}}]_{k,i} \Big)^2 \nonumber \\
&\leq&  \Big(\sum_k [\sqrt{\mathbf{F}}]_{i,k} [\sqrt{\mathbf{F}}]_{k,i}\Big) 
\Big(\sum_k [\sqrt{\mathbf{F}^{-1}}]_{i,k} [\sqrt{\mathbf{F}^{-1}}]_{k,i}\Big),  \nonumber
\eeq
we obtain 
\be
1 \leq \mathbf{F}_{i,i} [\mathbf{F}^{-1}]_{i,i} \qquad \forall i, \nonumber
\ee
and recover Eq.~(\ref{Ineq:Fii}). 
The equality sign is saturated if and only if
$\mathbf{F} = c \Eins$, where $c$ is a positive real number.

%%%%%%%%%%%%%%%%%%%%%%%%%%%%%%%%%%%%%%%%%%%%%%%%%%%%%%%%%%%
%%%%%%%%%%%%%%%%%%%%%%%%%%%%%%%%%%%%%%%%%%%%%%%%%%%%%%%%%%%
%%%%%%%%%%%%%%%%%%%%%%%%%%%%%%%%%%%%%%%%%%%%%%%%%%%%%%%%%%%

\section{Upper bound to ${F}_{i,i}$}

The following inequality holds
\be \label{Fii_ineq}
{F}_{i,i} \leq F_Q^{(i)} \big[ \hat{\rho}(\vect{\lambda}) \big],
\ee
where $F_Q^{(i)} \big[ \hat{\rho}(\vect{\theta}) \big]= [F_Q]_{i,i}$ is the diagonal element of the quantum Fisher information matrix,
\be \label{FQ}
F_Q^{(i)} \big[ \hat{\rho}(\vect{\theta}) \big]  = \tr[ \hat{\rho}(\vect{\lambda}) \hat{L}_i^2]
\ee
and $\hat{L}_i$ is the symmetric logarithmic derivative, satisfying 
\be \label{SLD}
\frac{\partial \hat{\rho}(\vect{\lambda})}{\partial \lambda_i} = \frac{ \hat{L}_i \hat{\rho}(\vect{\lambda}) + \hat{\rho}(\vect{\lambda}) \hat{L}_i}{2}.
\ee
The demonstration of the inequality (\ref{Fii_ineq}) follows from Ref.~\cite{BraunsteinPRL1994}
(see also the review \cite{PezzeVarenna}).
Using Eqs.~(7) and (\ref{SLD}), we have 
\be
\mathbf{F}_{i,i} =\sum_x \frac{1}{p(x \vert \vect{\lambda})}
\bigg( \frac{\partial p(x \vert \vect{\lambda}) }{\partial \lambda_i} \bigg)^2 
= \sum_x \frac{ \Re( \tr[\hat{\rho}(\vect{\lambda}) \hat{L}_i \hat{\Pi}_x] )^2 }{ \tr[\hat{\rho}(\vect{\lambda}) \hat{\Pi}_x)]}, \nonumber
\ee
where $\Re(y)$ is the real part of $y$. 
We then use the following chain of inequalities valid for all values of $x$:
\beq
\Re\big( \tr[\hat{\rho}(\vect{\lambda}) \hat{L}_i \hat{\Pi}_x] \big)^2 
&\leq& 
\big\vert \tr[\hat{\rho}(\vect{\lambda}) \hat{L}_i \hat{\Pi}_x] \big\vert^2 \nonumber \\
&\leq&
\tr\big[ \hat{\rho}(\vect{\lambda}) \hat{\Pi}_x \big] 
\tr\big[ \hat{\Pi}_x \hat{L}_i \hat{\rho}(\vect{\lambda}) \hat{L}_i \big], \nonumber
\eeq
the first inequality due to $\Re(y)^2 \leq |y|^2$ and the second is due to Caushy-Schwarz.
We thus obtain (for all values of $x$)
\be
\frac{ \Re( \tr[\hat{\rho}(\vect{\lambda}) \hat{L}_i \hat{\Pi}_x] )^2 }{ \tr[\hat{\rho}(\vect{\lambda}) \hat{\Pi}_x)]}
\leq \tr\big[ \hat{\Pi}_x \hat{L}_i \hat{\rho}(\vect{\lambda}) \hat{L}_i \big].
\ee
Summing over $x$, we thus recover Eq.~(\ref{Fii_ineq}) with $F_Q^{(i)} \big[ \hat{\rho}(\vect{\lambda}) \big]$
given in Eq.~(\ref{FQ}). 
The equality sign can be saturated by taking a projective measurement $\hat{\Pi}_x$ on the eigenstates of 
$\hat{L}_i$.

To find an explicit expression for $F_Q^{(i)} \big[ \hat{\rho}(\vect{\lambda}) \big]$, 
let us write $\hat{\rho}(\vect{\lambda})$ in diagonal form 
$\hat{\rho}(\vect{\lambda}) = \sum_k p_k \vert k \rangle \langle k \vert$, with $p_k \geq 0$, 
$\{ \vert k \rangle \}$ is a complete basis ($\sum_k \vert k \rangle \langle k \vert = \Eins$) and $\sum_k p_k=1$. 
We have
\be
F_Q^{(i)} \big[ \hat{\rho}(\vect{\lambda}) \big]  = 2 \sum_{\substack{ k,k' \\ p_k+p_{k'} \neq 0 } }
\frac{ (p_k -p_{k'})^2 }{ p_k +p_{k'} } 
\big\vert \langle k \vert \hat{H}_i \vert k' \rangle \big\vert^2, \nonumber
\ee
where 
\be
\hat{H}_i \equiv i \bigg( \frac{\partial \hat{U}(\vect{\lambda})}{\partial \lambda_i} \bigg) \hat{U}^{-1}(\vect{\lambda}).
\ee
In particular, 
\be
F_Q^{(i)} \big[ \hat{\rho}(\vect{\lambda}) \big]  \leq 4 \big( \Delta \hat{G}_i \big)^2.
\ee
The quantity $ [F_Q]_{i,i}$ has the physical meaning 
of a single-parameter quantum Fisher information 
with respect to transformations $ \hat{U}(\vect{\lambda})$ where all the parameters are fixed except $\lambda_i$
(in other words $\frac{\partial \hat{U}(\vect{\lambda})}{\partial \lambda_j}=0$ for $i\neq j$).

%%%%%%%%%%%%%%%%%%%%%%%%%%%%%%%%%%%%%%%%%%%%%%%%%%%%%%%%%%%
%%%%%%%%%%%%%%%%%%%%%%%%%%%%%%%%%%%%%%%%%%%%%%%%%%%%%%%%%%%
%%%%%%%%%%%%%%%%%%%%%%%%%%%%%%%%%%%%%%%%%%%%%%%%%%%%%%%%%%%

\section{Relation to qudit entanglement}

The relation between $ [F_Q]_{i,i}$ and qudit entanglement is a generalisation of 
the criteria of useful entanglement discussed in Ref.~\cite{PezzePRL2009}
(see also \cite{GiovannettiPRL2006} and the review \cite{PezzeVarenna}).
We recall that a state $\varrho_{0}$ of $N$ qudits is said to be qudit-separable if can be written as 
\be \label{Eq:separable}
\hat{\varrho}_{\rm sep} = \sum_k p_k \, 
\vert \psi_{k,1} \rangle \langle \psi_{k,1} \vert \otimes 
\vert \psi_{k,2} \rangle \langle \psi_{k,2} \vert \otimes ...
\vert \psi_{k,N} \rangle \langle \psi_{k,N} \vert, \nonumber
\ee
where $\vert \psi_{k,n} \rangle$ $(n=1,\cdots,N)$ is a single qudit state. 
We assume here that the Hamiltonian $\hat{G}_i$ is local in the qudits
(i.e. it can be written as $\hat{G}_i = \sum_{n=1}^N \hat{g}_i^{(n)}$, where $\hat{g}_i^{(n)}$
acts on the $n$th qudit), and, for simplicity, we take $\hat{g}_i^{(n)} = \hat{g}_i$  
for all $n=1,\cdots,N$.
Under these conditions we find 
\be \label{FQsep}
F_Q^{(i)}  \leq N (g_{i, \max} - g_{i, \min})^2,
\ee
where $g_{i, \max}$ and $g_{i, \min}$ are the maximum and minimum eigenvalues of $\hat g_i$.
Taking into account Eq.~(\ref{Fii_ineq}), we obtain that the inequality
\be
{F}_{i,i} \leq N (g_{i, \max} - g_{i, \min})^2
\ee
holds for all separable qudit states and all possible POVMs.
A violation of this inequality signals qudit entanglement. 

The demonstration of Eq.~(\ref{FQsep}) uses general properties of the single-parameter
quantum Fisher information \cite{PezzeVarenna}):
\begin{itemize}
\item the convexity of $F_Q^{(i)}$:
\be
F_Q^{(i)} \big[ \hat{\rho}_{\rm sep} \big] \leq 
\sum_k p_k F_Q^{(i)} \big[ \vert \psi_{1,k} \rangle \otimes ... \otimes \vert \psi_{N,k} \rangle \big]; \nonumber \\
\ee
\item the additivity of $F_Q^{(i)}$:
\be
F_Q^{(i)} \big[ \vert \psi_{1,k} \rangle \otimes ... \otimes \vert \psi_{N,k} \rangle \big] = 
\sum_{n=1}^N F_Q^{(i)} \big[ \vert \psi_{n,k} \rangle \big]; \nonumber \\
\ee
\item the bound 
\be
F_Q^{(i)} \big[ \vert \psi_{n,k} \rangle \big] \leq 4 \big( \Delta \hat{g}_i \big)^2_{\vert \psi_{n,k} \rangle}; \nonumber
\ee
\item the inequality 
\be
4\big( \Delta \hat{g}_i \big)^2_{\vert \psi_{n,k}\rangle} \leq \big( g_{i, \rm max} -  g_{i, \rm min} \big)^2, \nonumber
\ee
which holds for every $n$, i.e. for every single-qudit state.
\end{itemize}
Putting all these results together,
we arrive at Eq.~(\ref{FQsep}).

\section{The estimator for the adaptive protocol}

The chosen estimator is essentially a Maximum likelihood estimator, with likelihood function $\mathcal{L}({\bm \phi})$:
\begin{equation}
\mathcal{L}({\bm \phi}) = \log \left[ p^{(\alpha)}({\bm \phi}) \prod_{k=1}^{k_{\mathrm{max}}} p(k \vert {\bm \phi})^{n_{k}}\right].
\end{equation}
Here, $p^{(\alpha)}({\bm \phi})$ represents the knowledge on the parameters, $p(k \vert {\bm \phi})$ is the conditional probability of outcome $k$, and $n_{k}$ is the number of occurrence of outcome $k$. At each step, the distribution $p^{(\alpha)}({\bm \phi})$ is a Gaussian distribution:
\begin{equation}
p^{(\alpha)}({\bm \phi}) = \prod_{i=1}^{2} \frac{1}{\sqrt{2 \pi \sigma_{i}^{(\alpha)\,2}}} e^{-\frac{(\phi_{1}-\phi_{i}^{(\alpha)})^{2}}{(2 \sigma_{i}^{(\alpha)\,2})}},
\end{equation}
being $\phi_{i}^{(\alpha)}$ and $\sigma^{(\alpha)}_{i}$ the estimated value and the error on the parameter $i$ obtained at step $\alpha-1$ respectively.

%%%%%%%%%%%%%%%%%%%%%%%%%%%%%%%%%%%%%%%%%%%%%%%%%%%%%%%%%%%%%%%%%%%%%%%%%%%%%%%%%%%%%%%%%%%%%%%%%%%%%%%%%%%%%%%

%
\begin{figure*}[ht!]
\centering
\includegraphics[width=0.9\textwidth]{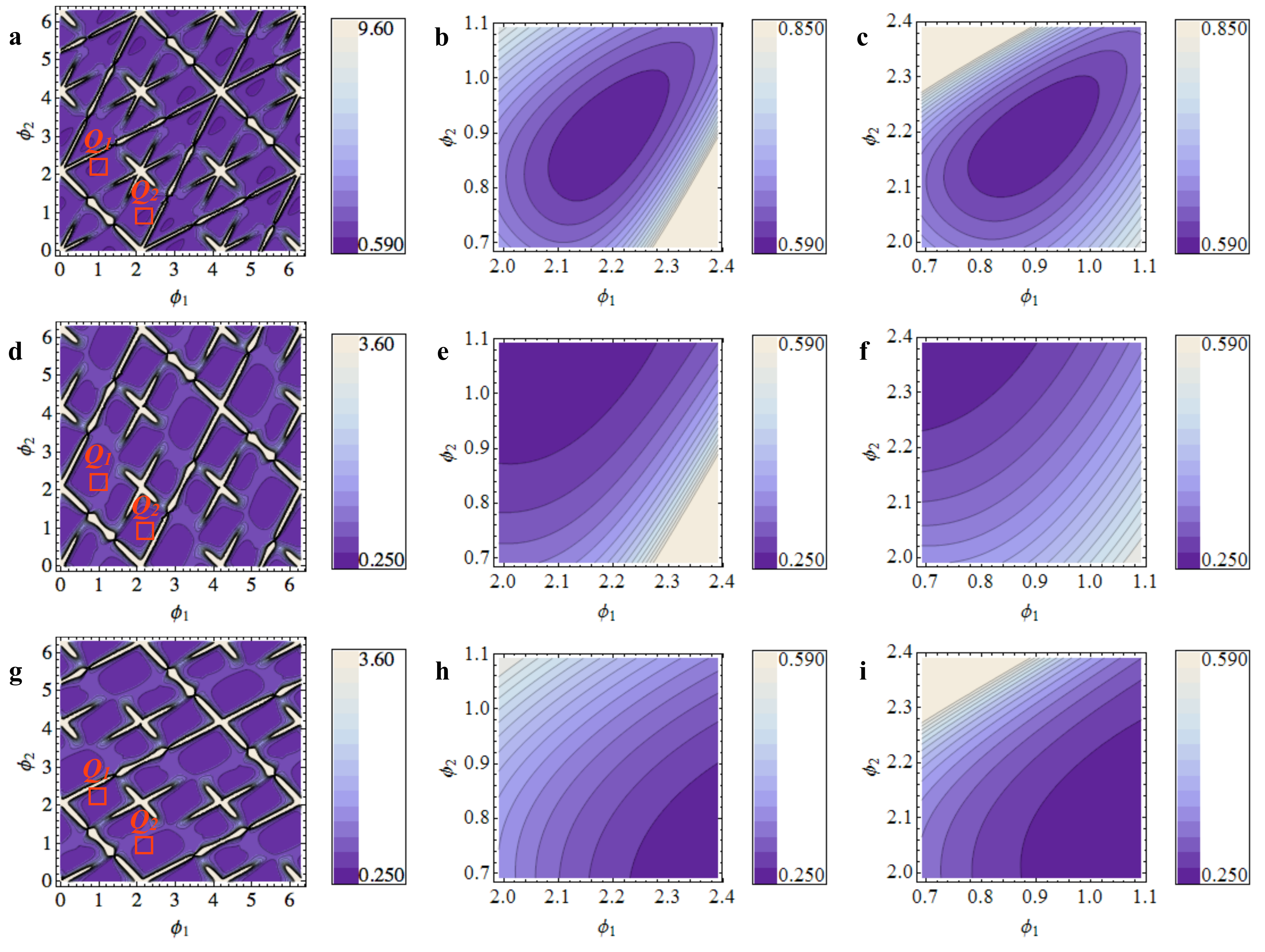}
\caption{Phase sensitivity of the three-mode balanced MZI with $\vert 1,1,1 \rangle$ probe state and photon-number measurement. \textbf{a}, Contour plot of $\mathrm{Tr}[\mathbf{F}^{-1}]$ as a function of $\phi_{1}$ and $\phi_{2}$. \textbf{b-c}, Regions of $\mathrm{Tr}[\mathbf{F}^{-1}]$ around the values of the phases which minimize $\mathrm{Tr}[\mathbf{F}^{-1}]$. \textbf{d}, working point $Q_{1}$ and \textbf{c}, working point $Q_{2}$.
\textbf{d}, Contour plot of $(\mathbf{F}^{-1})_{1,1}$ as a function of $\phi_{1}$ and $\phi_{2}$. \textbf{e-f}, Regions of $(\mathbf{F}^{-1})_{1,1}$ around the values of the phases which minimize $\mathrm{Tr}[\mathbf{F}^{-1}]$. \textbf{e}, working point $Q_{1}$ and \textbf{f}, working point $Q_{2}$. \textbf{g}, Contour plot of $(\mathbf{F}^{-1})_{2,2}$ as a function of $\phi_{1}$ and $\phi_{2}$. \textbf{h-i}, Regions of $(\mathbf{F}^{-1})_{2,2}$ around the values of the phases which minimize $\mathrm{Tr}[\mathbf{F}^{-1}]$. \textbf{h}, working point $Q_{1}$ and \textbf{i}, working point $Q_{2}$.}
\label{fig:CFI_SI}
\end{figure*}
\begin{figure}[t!]
\centering
\includegraphics[width=0.49\textwidth]{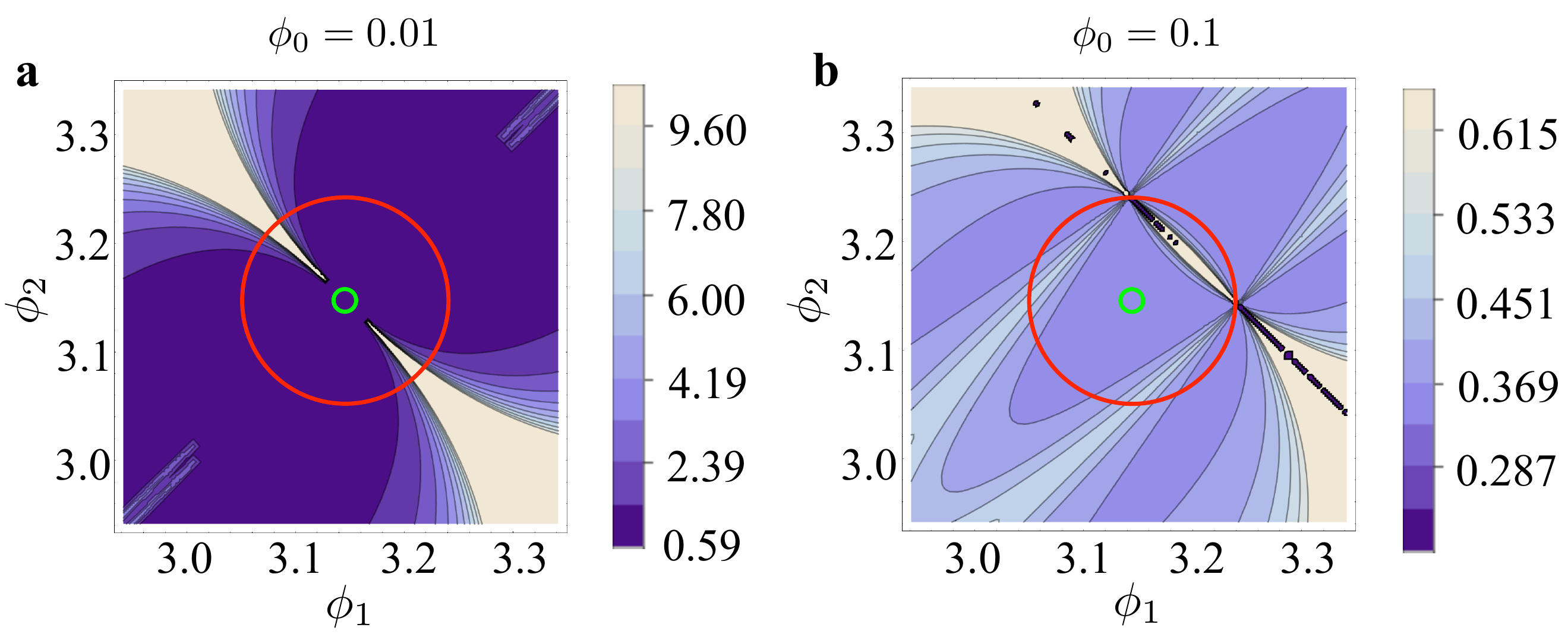}
\caption{Analysis of stability for the four-mode interferometer with $\vert 1,1,1,1 \rangle$ probe state and photon-number measurement. To perform a two-step adaptive protocol for the estimation of two unknown phases with the highest precision, we need that in a neighbourhood of radius $\delta$ centered in the working point $O_{1}$, the quantity Tr$[\mathbf{F}^{-1}]$ has no singularities.
{\bf a-b}, Contour plot of $\mathrm{Tr}[\mathbf{F}^{-1}]$ around the working point $O_{1}=[\pi,\pi]$ for {\bf a} $\phi_{0}=0.01$ and {\bf b} $\phi_{0}=0.1$. Green circles: regions with $\delta = 0.01$ around $O_{1}$. Red circles: regions with $\delta = 0.1$ around $O_{1}$. We note that increasing $\phi_{0}$ the singularities move away from the neighbourhood, increasing the stability of the estimation protocol.}
\label{fig:Stability}
\end{figure}